\documentclass[prd,showpacs,amsmath,showkeys,twocolumn,floatfix]{revtex4} 
\usepackage{epsfig,dcolumn}
\usepackage{graphicx}
\usepackage[usenames]{color}
\usepackage{graphicx}
\usepackage{bm}

\def\x{{\bf x}}
\def\y{{\bf y}}

\def\k{{\bf k}}
\def\q{{\bf q}}
\def\p{{\bf p}}

\def\A{{\bf A}}

\def\r{{\bf r}}
\def\a{\alpha}

\def\hk{\hat{\bf k}}

\def\l{{\bf l}}

\def\lsim{\mathrel{\rlap{\lower4pt\hbox{\hskip1pt$\sim$}}
    \raise1pt\hbox{$<$}}}
\def\gsim{\mathrel{\rlap{\lower4pt\hbox{\hskip1pt$\sim$}}
    \raise1pt\hbox{$>$}}}
\begin{document}


\title{ Gluelump spectrum from Coulomb gauge QCD} 

\author{ Peng Guo and Adam P. Szczepaniak}
\affiliation{ Physics Department and Nuclear Theory Center \\
Indiana University, Bloomington, IN 47405, USA. }

\author{ Giuseppe Galat\`a, Andrea Vassallo and  Elena Santopinto \\
I.N.F.N. and Dipartimento di Fisica, \\
via Dodecaneso 33, Genova, I-16146, Italy. }

\date{\today}

\begin{abstract}
We compute the energy spectrum of gluelumps defined as gluonic excitations bound to a localized, static octet source. We are able to reproduce the nontrivial ordering of the spin-parity levels
 and show how this is related to the non-abelian part of the Coulomb interaction between color charges. 
   \end{abstract}

\pacs{11.10Ef, 12.38.Mk, 12.40.-y, 12.38.Lg}

\maketitle

\section{Introduction} 
It is widely accepted that low energy gluons are responsible for the distinctive features of QCD such as confinement and chiral symmetry breaking. Due to strong interactions, precision calculations of gluonic excitations are sparse and experimental evidence for gluonic excitations is still murky.  This is because dynamical quarks and open channels need to be considered before comparing with experimental data.  
 For the purpose of getting insights into the non-perturbative features of QCD it is however  possible to design idealized systems that are sensitive to the pure Yang-Mills sector and to bypass many of the complications of the full QCD. A gluelump is an example of such an idealized state. It is a state of the gluon field bound to a static, localized octet source which can be constructed, for example as the quark-antiquark pair placed at an origin. Since the quark and the antiquark are at zero relative separation the system is rotationally symmetric, invariant under parity and charge conjugation. Thus gluelump states can be classified by the same $J^{PC}$ quantum numbers as ordinary mesons. 
The spectrum of gluelumps have been obtained from lattice gauge simulations and shows an unusual ordering of the various $J^{PC}$ levels~\cite{Foster:1998wu,Bali:2003jq}. The ground state has $J^{PC} = 1^{+-}$ and the first excited state has $J^{PC} = 1^{--}$. If the gluelump is to be interpreted as a state of a single (effective, quasi) gluon bound to a static source, then one would expect the $1^{--}$ state to be below $1^{+-}$, since the former corresponds to a single gluon in an $s$-wave orbital and the latter to a gluon in a $p$-wave orbital. The following two levels have $J^{PC}=2^{--}$ and $2^{+-}$ respectively and similarly correspond to a reverse ordering of the $p$-wave orbital ($2^{+-}$) and the $d$-wave orbital ($2^{--}$). It is clear that a simple constituent model of the gluon bound to the $Q{\bar Q}$ with a  central potential~\cite{Horn:1977rq} will not reproduce the lattice spectrum, since  the centrifugal barrier orders the spectrum  according to the orbital angular momentum. Another popular model for low lying gluonic excitations is based on the strong-coupling limit
  and considers gluons as excitations of the chromoelectric flux tube~\cite{Isgur:1984bm}. To the best of our knowledge the model has not been used to compute the gluelump spectrum but it does reproduce the ordering of the few lowest levels of the excited adiabatic potentials between the $Q{\bar Q}$ sources~\cite{Swanson:1998kx,Allen:1998wp, Buisseret:2006wc,Buisseret:2007ed}. The adiabatic potentials  give gluon energies for finite separation between the quark and the antiquark and reduce (modulo a zero point energy) to the gluelump spectrum in the limit of vanishing $Q{\bar Q}$ separation.  The correct ordering of the adiabatic potential in the flux tube model is  however somewhat artificial as it relies on a postulated intrinsic charge conjugation to be assigned to the flux tube. Furthermore,  while there is a clear evidence that at large separation between the $Q{\bar Q}$ sources the flux tube develops~\cite{Juge:2002br} and  gluonic excitations are effectively those of the Nambu-Goto string~\cite{Allen:1998wp}, on the contrary it is not clear if the flux tube picture is correct when the $Q{\bar Q}$ separation is taken to zero. In fact, the splitting of adiabatic potentials from lattice simulations  for small $Q{\bar Q}$ separations may be more constituent-like rather than string-like~\cite{Szczepaniak:2006nx}. The gluelump spectrum may therefore be used to  discriminate between alternative pictures of the gluonic excitations far from the adiabatic limit. Another reason for why it is important to pin down the characteristics of gluonic excitations for $Q{\bar Q}$ separation smaller than $\sim 1\mbox{ fm}$  is because of their relevance to the dynamics of heavy hybrid quarkonia. These are physical states and there already might be experimental evidence for such states~\cite{Swanson:2006st}. Finally mixing between the $Q{\bar Q}$ and the hybrid component of a heavy  quarkonium is responsible for relativistic corrections to the static $Q{\bar Q}$ potential which is relevant to normal quarkonium spectroscopy~\cite{Eichten:1980mw,Barchielli:1988zp,Brambilla:1993zw,Szczepaniak:1996tk}. 

In this paper we examine the  gluelump spectrum in the variational approach based on the Coulomb gauge QCD. A variational vacuum wave functional constrains single gluon properties and in particular leads to an effective gluon mass. This enables truncation of the gluon Fock space and makes it possible to  identify low lying gluelump states with the bound states of a single quasi-gluon. In Section II we present the details of the Coulomb gauge description of the gluelump spectrum and give the numerical results. Summary and outlook are summarized in Section III. 

\section{ The gluelump spectrum} 

The variational approach to Coulomb gauge QCD ~\cite{Christ:1980ku}  has recently  been studied in ~\cite{Swift:1988za,Zwanziger:1995cv,Szczepaniak:1996gb,Cucchieri:1996ja,Szczepaniak:2001rg,Szczepaniak:2003ve,Feuchter:2004mk,Reinhardt:2004mm}. Here we will only briefly  summarize  the method and give  the relevant equations for the study of gluelump spectrum.
The Coulomb  gauge eliminates unphysical degrees of freedom, provided the domain of transverse field variables, $\A^a(\x)$, ($\bm{\nabla} \cdot \A^a(\x) = 0$) is restricted to the fundamental modular region (FMR). The fundamental modular region is a subset of the Gribov region~\cite{Zwanziger:1995cv,vanBaal:1997gu} defined as the set of fields for which the determinant of the  Faddeev-Poppov (FP) operator  is positive and the FMR is a set of unique field representatives lying on a gauge orbit and satisfying the coulomb gauge condition.
  Restriction to the fundamental modular region is hard to implement, however it was argued in~\cite{Zwanziger:2003cf} that the bulk contribution to the functional integrals over the field variables comes from the common boundary of the FMR  and the Gribov region, and restriction to the Gribov region can be  implemented within  the variational approach. Within this approach the vacuum wave functional is chosen to be of the form 
\begin{equation}
\Psi[\A] = {\cal N}  \exp\left( - {1\over 2} \int d\x \A^a(\x) \omega(\x - \y) \A^a(\y) \right), \label{psi}
\end{equation} 
with the gap function $\omega$ determined by minimizing the expectation value of the Coulomb gauge Hamiltonian $H = H[\A, -i\delta/\delta \A]$
\begin{equation} 
{ {\delta} \over {\delta \omega}} \int D\A {\cal J} \Psi[\A] H\left[ \A, -i{{\delta} \over {\delta \A}} \right] 
\Psi[\A]  = 0 \label{gap} 
\end{equation} 
Here ${\cal J} = \mbox{Det}\bm\left[-{\nabla}\cdot {\bf D}[\A]\right]$ is the determinant of the Faddeev-Popov operator; ${\bf D} = {\bf D}_{ab} = \delta_{ab} \bm{\nabla}  + g f_{acb} {\bf A}^c$  is the covariant  derivative in the adjoint representation. The Coulomb gauge Hamiltonian $H$ contains direct interaction between color charge densities, $\rho^a(\x)  = \psi^{\dag}(\x) T^a \psi(\x) + f_{abc} \bm{\Pi}^b(\x) \cdot \bf{A}^c(\x)$, 
\begin{equation}
H_C = {{g^2}\over 2} \int d\x d\y \rho^a(\x) K(\x,a;\y,b) \rho^b(\y)
\end{equation} 
where
\begin{equation}
K(\x,a;\y,b) = \left\{ \left[ \bm{\nabla} \cdot {\bf D} \right]^{-1} (-\bm{\nabla}^2) \left[ \bm{\nabla} \cdot {\bf D} \right]^{-1}\right\} _{\x,a;\y,b} \label{kernel}
\end{equation} 
The inverse of the two covariant derivatives, when expanded in power series in the coupling constant, $g$, produces an (infinite) series of interactions that couple an arbitrary number of transverse gluons to the quark and/or gluon sources via the bare Coulomb potential $-1/\bm{\nabla}^2$. 
When these gluons are integrated over the vacuum wave functional of Eq.~(\ref{psi}), they dress the bare Coulomb potential and it results in an effective interaction which  is approximately  linear for large separations between the sources~\cite{Cucchieri:1996ja,Szczepaniak:2001rg,Feuchter:2004mk}. This effective Coulomb potential is defined as 
\begin{equation}
V_{CL}(p)\delta_{ab}  = -\int d\x e^{i\p\cdot\x} \langle K(\x,a;{\bf 0},b) \rangle \label{effective} 
\end{equation}
with 
\begin{equation}
\langle K(\x,a;{\bf 0},b) \rangle   \equiv \int D\A {\cal J}  e^{i\k\cdot \x} K(\x,a;{\bf 0},b) |\Psi[\A]|^2 .
\end{equation} 
The subscript $CL$ means that $V_{CL}$ is expected to have a short distance, Coulomb-like component and a long-distance, linear part. The Dyson series for the effective interaction, $V_{CL}$  is divergent in the ultraviolet and the bare coupling is renormalized by  fixing  the slope of the potential to agree with the lattice data for the Wilson loop.  There is, however, a difference between the renormalization group equations for the coupling obtained in the  variational approach and that of the QCD beta-function. This is to be expected since the variational approximation includes only two-body correlations ( just like BCS) and, for example does not take into account propagation of transverse gluons~\cite{Swift:1988za,Szczepaniak:2001rg}. Furthermore, even  for exact vacuum the expectation value of $H_C$ is not the same at the energy of the $Q{\bar Q}$ state~\cite{Greensite:2003xf}. This is because the  state of a $Q{\bar Q}$ pair added to the vacuum is not the same as the true  eigenstate  of the QCD Hamiltonian in presence of the $Q{\bar Q}$ pair. All this lead to an ambiguity in fixing the  long range part of the $Q{\bar Q}$ potential. Fortunately we expect that this will not play an important role for small, and in the case of gluelumps, vanishing $Q{\bar Q}$ separations, and in general for  low energy gluonic   excitations that have wave functions confined to a region of  $\sim 1-2\mbox{ fm}$ that  are not  
 too sensitive to the asymptotic, long range behavior of the confining potential. Finally, ambiguities in the renormalization group equations of the variational methods, have so far prevented from obtaining a unique solution for $\omega$ by using the gap equation~\cite{Szczepaniak:2003ve,Schleifenbaum:2006bq,Epple:2006hv}. In particular it is expected that at low momentum, $\omega(p)$, (the Fourier  transform of $\omega(\x - \y)$) has momentum dependence identical to that of the expectation value of the curvature, $\chi(p)$  defined through the determinant of the Faddeev-Popov operator, 
\begin{equation} 
\omega(p)\delta_{ab}  \to -{1\over 2}   \int d\x  e^{i\p\cdot \x} \langle {{\delta^2 \ln {\cal J} } \over {\delta \A^a(\x) \A^b(0) }}  \rangle 
\end{equation}  
while at large momenta, $\omega(p) \sim p$. There are indications from lattice computations that 
  $\omega(p)$ is different from zero as $p \to 0$ but the lattice  results cannot  resolve between 
   $\omega(p \to 0)$ finite and $\omega(p \to 0 ) \to \infty$~\cite{Cucchieri:2006za,Langfeld:2004qs}. In our analysis we will thus study 
 how the gluelump spectrum depends on the gluon gap function $\omega$. 
     
The gap function $\omega$ is not the same as the single quasi-gluon energy, $E(p)$, although they are 
 closely related. For $\omega(p)$ satisfying the gap equation, Eq.~(\ref{gap}), the single particle energy is given by 
\begin{equation}
E(p) = \omega(p) + \Sigma(p) \label{spe}
\end{equation}
with the self-energy given by 
\begin{equation} 
\Sigma(p) = -{{N_C}\over 4}  \int {{d\k} \over{(2\pi)^3}} 
 V_{CL}(|\p - \k|) \left[ 1 + (\hat\p\cdot\hat \k)^2 \right] {{\omega(p)} \over {\omega(k)}} . \label{self}  
\end{equation} 
The long range nature of the Coulomb kernel $K$, Eq.~(\ref{kernel}), leads to strong IR enhancement in 
$V_{CL}(p)$, $V_{CL}(p\to 0) \to \infty$, which in turn makes the integral in Eq.~(\ref{self})  divergent for $\p \sim \k$. This is a signature of color confinement in the mean field which leads to infinite energies for  colored states. The total energy of a color singlet state, in particular gluelumps, is however  finite, as will be seen  below,  with all theIR divergences canceling among various contributions. 

\subsection{Fitting the potential} 

\begin{figure}
\includegraphics[width=2.5in,angle=270]{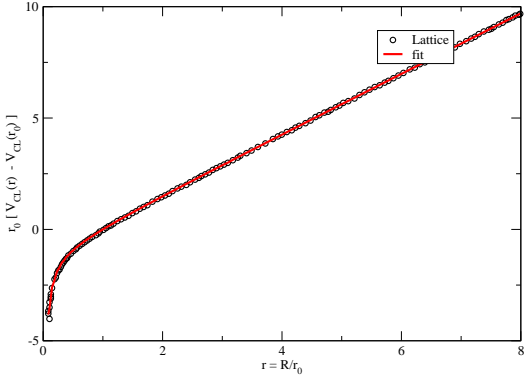}
\caption{\label{fit-with-log} Lattice data and fit results. The errors on the lattice data are taken to be $\delta \epsilon = 0.05$ and the fit gives $\chi^2/p.d.f =1.85$}  \end{figure}

The solution of the Dyson equation for the effective interaction Eq. ~(\ref{effective}) can be written in terms of the square of the expectation value of the inverse Faddeev-Popov operator,$d(p)$, 
\begin{equation}
{{d(p)}\over {p^2}} \delta_{ab} = -\int d\x e^{i\p\cdot \x} \langle {g\over {\bm{\nabla}\cdot {\bf D}} } \rangle(\x,a;0,b) 
\end{equation}
and the Coulomb form factor, $f(p)$ which measures the difference between the square of the expectation value of the FP operator and  the expectation value of the square of the FP operator which defines the Coulomb kernel Eq.~(\ref{kernel}),
\begin{equation}
V_{CL}(p) = - {{d^2(p) f(p)} \over {p^2}} \label{f2d} .
\end{equation} 
As discussed  above, the mean field approximation, by neglecting transverse gluon loops, leads to a 
 high momentum behavior of $V_{CL}(p)$, which is different from that of QCD. For $p \to \infty$ 
$V_{CL}(p) \propto 1/[p^2\log^{3/2}(p)]$.  In the low momentum regime, to reproduce the linear confinement,
 $V_{CL}(p)$ should grow as $1/p^4$  when $p \to 0$.  Unfortunately, in the mean field approximation  
  common solutions of the Dyson equations for $d$, $f$ and $\omega$ do not quite reproduce such a strong IR enhancement of $V_{CL}$. In the so called angular approximation ~\cite{Szczepaniak:2001rg} it is possible to obtain a solution with both $d(p)$ and $f(p)$  divergent in the limit $p\to 0$ resulting in 
   $V_{CL}(p) \propto 1/p^{15/4}$. It was found in~\cite{Reinhardt:2004mm,Schleifenbaum:2006bq}, however, that without the angular approximation  
    the three Dyson equations for $d$, $f$  and $\omega$  do not have solutions such that both $d(p)$ and $f(p)$ are divergent in the IR (as $p \to 0$). The IR behavior of the solutions is controlled by the value of the  renormalized  coupling $g \to g(\mu)$. In the mean filed approximation it is found that there is 
      a critical coupling $g_c(\mu)$ for which $d(p)$ is IR divergent, while for all $g(\mu) < g_c(\mu)$ both $d(p)$ and $f(p)$ saturate in the IR.  These sub-critical solutions do not, in a strict sense lead to confinement, however, for all practical application $V_{CL}(r)$ (defined as a three-dimension Fourier transform of $V_{CL}(p)$) computed from these sub-critical $d$ and $f$ is very close to a linear function of $r$ for a $r \lsim  10\mbox{ fm}$.  We will thus approximate $V_{CL}(r)$ by a linear potential for $r$ in this range plus a short range, Coulomb piece modified by the loop corrections as discussed above. In momentum space,
      \begin{equation} 
 V_{CL}(p) = V_C(p) + V_L(p)
 \end{equation}
 with the Coulomb part, $V_C$, and the linear part, $V_L$, given by
\begin{eqnarray}
 V_C(p) & = &  -{{4\pi \alpha(p)} \over {p^2}} ,\;\;  \alpha(p) = {{4\pi Z}  \over { \beta^{3\over 2}  
\log^{3\over 2} \left( {{p^2} \over {\Lambda_{QCD}^2} } + c\right)} } 
  \nonumber \\
V_L(p) & =  & -{{8\pi b}  \over {p^4}} 
 \end{eqnarray}
The Fourier transform of $V_L$ which defines the coordinate space potential is ill-defined due to the strong IR singularity in $V_L$, and leads to an undetermined constant in $V_{CL}(r)= V_C(r) + V_L(r)$. In computations of $V_{CL}(r)$, including lattice, this constant is removed by computing a difference $V_{CL}(r) - V_{CL}(r_0)$ at some fixed $r_0$:
\begin{equation}
V_L(r) - V_L(r_0) = - \int {{d\p} \over {(2\pi)^3}} \left[ e^{i\p\cdot \r} - e^{i\p\cdot\r_0} \right] {{8\pi b} \over {p^4}}.
\end{equation} 
If we were to remove the logarithm in the Coulomb part $\alpha(p) \to \alpha = \mbox{ const.}$ then also the Coulomb part would have a simple position space representation
\begin{equation}
V_C(r) = - \int { {d\p}  \over {(2\pi)^3}} e^{i\p\cdot\r} {{4\pi \alpha} \over {p^2}} = -{\alpha \over r}.
\end{equation}
In the following, however,  we will keep the full expression for $\alpha(p)$ and compute the position space potential $V_C(r)$ numerically 
\begin{equation}
V_C(r) = - (4\pi) \int_0^\infty { {p^2 dp}  \over {(2\pi)^3}} j_0(rp)  {{4\pi \alpha(p) } \over {p^2}}.
\end{equation} 
We fix the four parameters,  $Z,c, \Lambda_{QCD},b$, by  fitting the position space potential 
 $V_{CL}(r) - V_{CL}(r_0)$ to the lattice data~\cite{Juge:1997nc} with $r_0 \sim 1/450\mbox{ MeV}$. 
  The result of the fit is shown in Fig.~\ref{fit-with-log}. Unfortunately the low-r lattice data does not go to small enough distances to fix the Coulomb potential uniquely. Thus in the fit we fixed $\Lambda_{QCD} = 250\mbox{ MeV}$,
 and then  obtain 
 \begin{equation}
 b = 0.204 \mbox{ GeV}^2,\;\; Z = 5.94,\;\;c = 40.68..
 \end{equation}

\subsection{Basis of Gluelump states} 


The mean field vacuum leads to a basis of quasi-particle states which can be used to express any state of the Yang-Mills field. The quasi-particle creation and annihilation operators are defined in terms of the field variables
\begin{eqnarray}
\A^a(\x) & = &  \sum_{\lambda} \int {{d\k} \over {(2\pi)^3}} {1\over {\sqrt{2\omega(k)}}} [ \a(\k,\lambda,a)\bm{\epsilon}(\k,\lambda) \nonumber \\ 
 & + &  \a^{\dag}(-\k,\lambda,a) \bm{\epsilon}^*(-\k,\lambda)] e^{i\k\cdot \x} \nonumber \\
\bm{\Pi}^a(\x) & = &  -i\sum_{\lambda} \int {{d\k} \over {(2\pi)^3}} \sqrt{{\omega(k)}\over 2} [ \a(\k,\lambda,a)\bm{\epsilon}(\k,\lambda) \nonumber \\
 & - &  \a^{\dag}(-\k,\lambda,a) \bm{\epsilon}^*(-\k,\lambda)] e^{i\k\cdot \x}, \nonumber \\
\end{eqnarray}
where $\bm{\Pi}$ is the variable canonically conjugated to $\A$ (in the Schr\"odinger  representation used in Sec.II $\bm{\Pi}^a(\x) = -i\delta/\delta\A^a(\x)$) and $\epsilon(\k,\lambda)$ is the  
 wave function of a state with  helicity-$\lambda$, $\lambda=\pm 1$.  
With $\omega$ satisfying the gap equation, the Hamiltonian rewritten in terms of the particle operators 
 $\a,\a^{\dag}$ has  a diagonal one-body part, {\it i.e} it contains only the ladder operator 
 \begin{equation}
 K = \sum_{\lambda,a}\int {{d\k} \over {(2\pi)^3}} E(k) \a^{\dag}(\k,\lambda,a)\a(\k,\lambda,a),
\end{equation} 
 while operators of the form $\a \a$ and $\a^{\dag}\a^{\dag}$ are removed by the gap equation. 
 The single particle energy $E(k)$ is given by Eq.~(\ref{spe}). Since (after IR regularization)  the quasi-particles are massive,  it is  reasonable to expect that light gluelump states will contain a minimal number, which is one,  of quasi-particles constrained by various symmetries. The state of a single-quasigluon gluelump is given by 
\begin{eqnarray}
 | J^{P}M \rangle & = &  \sum_{\lambda\pm 1}  \int {{d\k } \over {(2\pi)^3}} 
Q^{\dag}_i   {\overline Q}^{\dag}_j 
 \a^{\dag}(\k,\lambda ,a) |0\rangle   \nonumber \\ 
  & \times &  {{T^a_{ij}} \over {\sqrt{C_F N_C}}}  \Psi^{J^P}_M(\k\lambda). \label{state}
\end{eqnarray} 
The $SU(N_C)$ generators, $T^a$ couples the $N_C$-quasi-gluons to the quark-antiquark state in the adjoint representation resulting in a color-singlet gluelump state ($C_F = (N_C^2-1)/2N_C$).  
 Quark spin and position quantum numbers are implicit in the quark ($Q^{\dag}$) and antiquark (${\overline Q}^{\dag}$)  creation operators. The orbital wave function describes spin-$1$ quasi-gluon with helicity $\lambda=\pm 1$ projected onto a state with total spin-$J$, projection-$M$ and parity-$P$ and it is given by 
\begin{equation} 
\Psi^{J^P}_M(\k\lambda) = \sqrt{ {2J+1} \over {4\pi} } \sum_{\lambda' =\pm 1} \chi^{J^P}_{\lambda\lambda'}  
    D^{J*}_{M,\lambda'}(\hk)   \psi^{J^P}(k).  
  \end{equation}
 Here $\chi_{\lambda\lambda'}$
 distinguishes between the TM and TE modes, $\chi_{\lambda\lambda'} = \delta_{\lambda\lambda'}/\sqrt{2}$ for TM (natural parity) and 
  $\chi_{\lambda\lambda'} = \lambda\delta_{\lambda\lambda'}/\sqrt{2}$ for TE (unnatural parity) modes, respectively,
 and $\psi^{J^P}(k)$ is the orbital wave function which has to be determined by  diagonalizing  the QCD Coulomb gauge Hamiltonian in this single quasi-gluon subspace.  It follows from the properties of the $D$-functions that the radial  wave functions 
are normalized by 
  \begin{equation}
\ \int {{dk k^2}\over {(2\pi)^3}}  \psi^{J^P} (k) \psi^{J'^{P'}}(k)= \delta_{JJ'} \delta_{PP'}.
  \end{equation}
 Finally  we note that all  gluelump states built from single quasi-gluon have negative charge conjugation. 
 
 \subsection{ Hamiltonian matrix elements } 
\begin{figure}
\includegraphics[width=2.5in,angle=0]{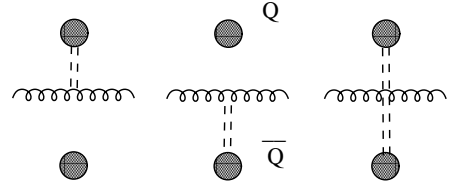}
\caption{\label{potential} Two -(left) and three- (right) body potential between the quasi-gluon and the $Q{\bar Q}$ pair. The static quark and the antiquark are represented by blobs. The dashed line represents the effective $V_{CL}$ potential (left) or the effective three-body interaction from Eq.~(\ref{3bkernel})(right). } \end{figure}

 Having constructed the basis which complies with the global symmetries of the Hamiltonian, we can compute the matrix elements in the single quasi-gluon subspace. Digitalization of the Hamiltonian matrix   
   leads to a one-dimensional Schr\"odinger equation for the radial wave functions, $\psi^{J^P}(k)$.
   \begin{eqnarray}
E^{J^P} \psi^{J^P}(p) &=  & \left[  \omega(p) + \Sigma(p)  +  2m _Q + \Sigma_{Q\bar Q} + V^8\right] \psi^{J^P}(p)  \nonumber \\
 &   + &    \int {{dk k^2}\over {(2\pi)^3}} V^{J^P}(p,k) \psi^{J^P}(k)  \label{Se}
 \end{eqnarray}
 Here $\omega + \Sigma$ is the quasi-gluon kinetic energy given in Eq.~(\ref{self}), $\Sigma_{Q\bar Q}$ is the sum of source (quark an antiquark) self energies, 
 \begin{equation}
 \Sigma_{Q\bar Q} = -C_F \int {{d\l} \over {(2\pi)^3}} V_{CL}(l), 
 \end{equation} 
 and $V^8$ is the potential energy of the octet $Q{\bar Q}$ pair at zero separation
 \begin{equation}
 V^8 = -{1\over {2N_C}} \int {{d\l} \over {(2\pi)^3}} V_{CL}(l). 
 \end{equation}
 The $Q{\bar Q}$ self energy and the $V^8$ are infinite. The infinities come both  from the 
    short distance part of the $V_C$  potential and from the long range part of the linear potential  
    $V_L$. Part of the UV divergence should be removed by the quark masses ($m_Q$)  and part needs to be subtracted because 
the $Q{\bar Q}$ pair at zero separation has infinite chromo-electrostatic energy. 
The IR divergence is a manifestation of the  long-range nature of the interaction and as long as the system is color-neutral, which is the case here,  should cancel with similar contributions from gluon-self energies and residual interactions.  In order to  remove the  infinite chromo-electrostatic energy the gluelump energy is defined as~\cite{Bali:2003jq}
    \begin{equation}
    E^{J^P}_G \equiv E^{J_P} - E_{Q\bar Q} - V^8_C + V^0_C,
    \end{equation}
   where $E_{ Q\bar Q}$ is the  total energy of the color-singlet,   static $Q{\bar Q}$ source 
   \begin{equation}
   E_{ Q\bar Q} = 2m_Q + \Sigma_{Q\bar Q} + V^0, 
   \end{equation}
   with 
   \begin{equation}
   V^0 = C_F \int {{d\l} \over {(2\pi)^3}} V_{CL}(l)
   \end{equation}
 being the potential energy of the singlet $Q{\bar Q}$ pair; $V^8_C$ and $V^0_C$ are the short distance parts of the octet and singlet $Q{\bar Q}$ potential energy, 
  \begin{eqnarray}
  V^8_C  & = &  -{1\over {2N_C}} \int {{d\l} \over {(2\pi)^3}} V_C(l) \nonumber \\
   V^0_C & = & C_F \int {{d\l} \over {(2\pi)^3}} V_C(l),  
   \end{eqnarray} 
  respectively.   Combining these in the  Sch\"odinger equation for the gluelump energy we obtain 
  \begin{eqnarray} 
 E^{J^P}_G  \psi^{J^P}(p) &  = & \left[  \omega(p) + \Sigma'(p) \right] \psi^{J^P}(p)  + 
    \nonumber \\
      & + &  \int {{dk k^2}\over {(2\pi)^3}} \left[V^{J^P}(p,k) + \Delta V^{J^P} (p,k) \right]   \psi^{J^P}(k)  \nonumber \\
  \label{Sef}
 \end{eqnarray}
where 
\begin{equation} 
\Sigma'(p) = -{{N_C}\over 2}  \int {{d\k} \over{(2\pi)^3}} 
 V_{CL}(|\p - \k|) \left[  {{ 1 + (\hat\p\cdot\hat \k)^2} \over 2} {{\omega(p)} \over {\omega(k)}}  - 1 \right] 
\end{equation} 
and 
\begin{equation} 
\Delta V^{J^P} = -N_C {{(2\pi)^3} \over {k^2}} \delta(k-p)  \int {{d\l} \over {(2\pi)^3}} V(l). 
\end{equation} 
The modified gluon self energy $\Sigma'$ is now finite in both IR and UV and, as will be seen below, 
$\Delta V$ removes the IR divergence from the matrix elements of the potential between the quasi-gluon and the quark  or the antiquark. The potential $V^{J^P}$  is a sum of two- and three-body interactions. 
 \begin{equation}
V^{J^P} = V^{J^P}_2 + V^{J^P}_3.
\end{equation} 
The two body interaction acts between the quasi-gluon and the quark or the antiquark while the three-body terms involve all three particles. Both terms originate from the Coulomb kernel of the QCD Hamiltonian and are depicted in Fig.~\ref{potential}. In the two body interaction all gluons emerging from the Coulomb kernel, Eq.~(\ref{kernel}), are integrated over the vacuum wave functional, producing the effective interaction $V_{CL}$ which acts as a potential between the quasi gluon and the quark or the antiquark. In the three-body term $V^{J^P}_3$, the gluon in the gluelump is contracted with one of the gluons in the Coulomb kernel, producing an effective interaction which involves all the three constituents, the gluon, the quark and the antiquark.
The two-body interaction, combined with  $\Delta V^{J^P}$, is given by 
\begin{widetext}
\begin{equation}
V^{J^P}_2(p,k)  = + {N_C\over 4} \left[ 
\sqrt{{{ \omega(k)}\over {\omega(p)}}} + \sqrt{{{\omega(p)}\over {\omega(k)}} }\right] 
4\pi  \sum_{l} 
 \left[ 1 - P(-1)^{l} \right] 
  {{2l+1}\over {2J+1}} \langle 11,l0|J1\rangle^2 
 \int d(\hat\p\cdot\hat\k)  V'_{CL}(\k -\p) P_l(\hat\p\cdot\hat\k) 
 \label{vps}
 \end{equation}
 \end{widetext}
where 
\begin{equation}
V'_{CL}(|\k-\p|) \equiv V_{CL}(|\k-\p|) - (2\pi)^3\delta(\k-\p) \int {{d\l} \over {(2\pi)^3}} V_L(l) 
\end{equation}
is IR finite. 
For natural parity states, $P = (-1)^J$, the two-body potential  reduces to 
\begin{widetext}
\begin{equation}
V^{J^P}_2(p,k)  = + {N_C\over 4} \left[ 
\sqrt{{{ \omega(k)}\over {\omega(p)}}} + \sqrt{{{\omega(p)}\over {\omega(k)}} }\right] 
{{4\pi} \over {2J+1}} \nonumber \\
 \int d (\hat\p\cdot\hat\k)  V'_{CL}(\k -\p) [(J+1) P_{J-1} (\hat\p\cdot\hat\k)  
  + J P_{J+1} (\hat\p\cdot\hat\k) ], \label{natural}
\end{equation}
\end{widetext}
and for unnatural parity states, $P =(-1)^{J+1}$, 
\begin{widetext}
\begin{equation}
V^{J^P}_2(p,k)  = + {N_C\over 4} \left[ 
\sqrt{{{ \omega(k)}\over {\omega(p)}}} + \sqrt{{{\omega(p)}\over {\omega(k)}} }\right] 
4\pi \nonumber \\
 \int d(\hat\p\cdot\hat\k) V'_{CL}(\k -\p)  P_J(\hat\p\cdot\hat\k) . \label{unnatural} 
\end{equation}
\end{widetext}
The three-body interaction is by itself IR finite and its matrix elements are given 
by~\cite{Szczepaniak:2005xi}
\begin{widetext} 
\begin{eqnarray}
&&V(p,q) = -{{N_C^2} \over 4}{{(4\pi)^2} \over {\sqrt{2\omega(p) 2\omega(q)}}} 
\sum_{l_q,l_p} 
  \left[ 1 - P(-1)^{l_q} \right] \langle l_q 0,11|J1\rangle 
\langle l_p 0,11|J1\rangle 
\langle l_q 0,10|J0\rangle \langle l_p 0,10|J0\rangle  \nonumber \\
& \times &  {{(2l_q+1) (2l_p+1) } \over {(2J+1)^2}}  \int {{dk k^2}\over {(2\pi)^3}}   \left[ 
  \int d(\hat\q\cdot\hat\k) d(\hat\p\cdot\hat\k) k^2 K(|\q + \k|,|\k|,|\k + \p|) P_{l_q}(\hat\q\cdot\hat\k) 
 P_{l_p}(\hat\p\cdot\hat\k)  \right],   \nonumber \\ \label{3b} 
\end{eqnarray} 
\end{widetext} 
where the three-body kernel is
\begin{eqnarray}
K(p,k,q) & = &  -{{d(p)d(k)d(q)} \over {p^2k^2q^2}} \nonumber \\
& \times &  \left[ f(p)d(p) + f(k)d(k) + f(q) d(q) \right]. \label{3bkernel}
\end{eqnarray}
Each of the three terms in square bracket  is given by a product of a $V_{CL}$  potential, Eq. ~(\ref{f2d}), in either $p$, or $k$ or $q$ multiplied by bare Coulomb potentials in the other two  momenta  corrected by the ghost form factors. In numerical computations we approximate these two modified coulomb potentials by $V_C$ times a scale factor, $\kappa$, which is fitted to the data. Thus, finally 
\begin{eqnarray} 
K(p,q,k) &= & \kappa \left[ V_{CL}(p) V_C(q) V_C(k) + V_C(p) V_{CL}(q) V_C(k) \right.  + \nonumber \\
&  + &  \left. V_C(p) V_C(q) V_{CL}(k)  \right]. \label{v3bk}
\end{eqnarray}

\subsection{Numerical results} 

Since the quasi-gluons have helicity $\pm 1$ 
the Wigner-$D$ functions in Eq.~(\ref{state}) restrict the spin of the low lying gluelumps, saturated by a single quasi-gluon, to $J \ge 1$. This is in agreement with lattice 
 computations~\cite{Foster:1998wu,Bali:2003jq}.  The two lowest $J$-states,  are $J^P = 1^-$ and $J^P=1^+$ . For the natural parity  the gluelump  state is a mixture of the $l=J-1=0$, $s$-wave, and $l=J+1=2$, $d$-wave, quasi-gluon state, which is explicit in the form of the two-body interaction in Eq.~(\ref{natural}). The unnatural parity state contains the quasi-gluon in the  $l=J=1$, $p$-wave state, {\it cf.} Eq.~(\ref{unnatural}). It is therefore expected that the two-body interaction alone would 
  lead to the energy of the natural parity, $J^P=1^-$, state to be lower than that of the unnatural parity, $J^P=1^+$, state.  The lattice computations, however find the $J^P=1^+$ state to have lower energy than the $J^P=1^-$ state. This is due to the three-body potential $V^{J^P}_3$. Indeed, for $J=1$ and positive parity, in Eq.~(\ref{3b}) the $[1-P(-1)^{l_q}]$ term implies $l_q =odd$ and the Clebsch-Gordon (CG) coefficient $\langle l_q0,10|J=1,0\rangle$ vanishes. For $J=1$ and negative parity, the product of all the four CG coefficients is positive. Thus the three-body interaction does not contribute to the $J^P=1^+$ state, while it increases the energy of the $J^P=1^-$ state. The three-body interaction can therefore change the ordering of the $1^-$ and $1^+$ state and push the $1^-$ state above the $1^+$ state. Similar inversion can take place for higher $J$ states. In general $V^{J^P}_3$ vanishes for {\it all} unnatural parity states. Thus, while for given $J$ 
    the two-body potential is expected to produce  the natural parity state with lower energy than the unnatural parity partner, the three-body interaction will add to the energy of the natural parity state, possibly pushing it above the unnatural parity one. This happens in the lattice data at least for the first  five  states. In fact $1^+$ is below $1^-$. The next state is the $2^-$ and is below the $2^+$ state, which is  followed by the $3^+$. The sixth state is the $0^{++}$ which, having both positive charge conjugation and $J=0$, in the quasi-gluon picture would require two gluon constituents. 
    
\begin{figure}
\includegraphics[width=2.5in,angle=270]{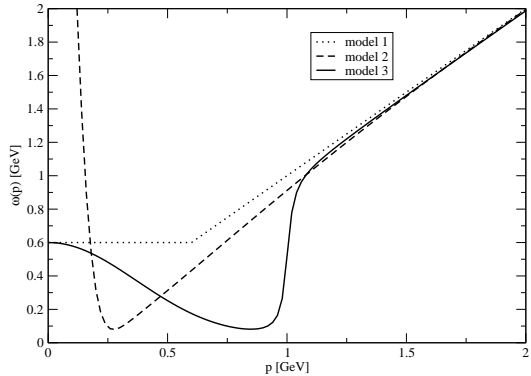}
\caption{\label{omega-fig} Three models for the gap function $\omega(p)$.}
 \end{figure}

\begin{figure}
\includegraphics[width=2.5in,angle=270]{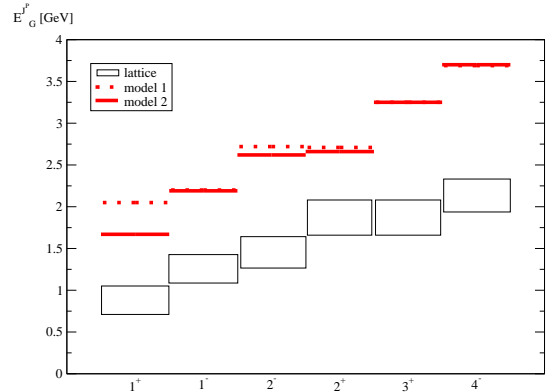}
\caption{\label{lattice_and_omega1and2} Gluelump spectrum from Eq.~(\ref{Sef}) with $\omega$ from models $1$ and $2$ shown in Fig.~\ref{omega-fig} and $\kappa = 0.6$ (Eq.~(\ref{v3bk})) compared with  results of lattice computations ~\cite{Bali:2003jq}. }
 \end{figure}

\begin{figure}
\includegraphics[width=2.5in,angle=270]{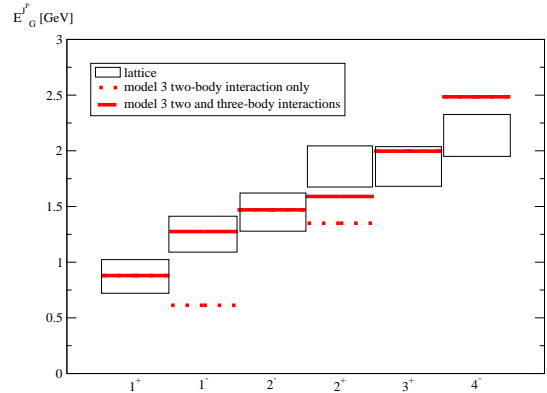}
\caption{\label{lattice_and_omega3}  Gluelump spectrum with $\omega(p)$ from model-$3$,  
 shown in Fig.~\ref{omega-fig}, and $\kappa = 0.4$ (Eq.~(\ref{v3bk})). The dashed lines correspond to the spectrum obtained with two-body interaction only and the solid lines correspond to the full calculation.  }
 \end{figure}

    For each $J^P$ gluelump state the Schr\"odinger equation, Eq.~(\ref{Sef}), is solved by  discretizing the momentum coordinate through a Gauss-Legendre grid. We examine sensitivity of  the computed spectrum to the form of the gap function $\omega$. In particular we first try two cases. One with 
     $\omega(k \to 0) = finite$ and the other with $\omega(k\to 0) \to \infty$. The former has been found as an approximate solution to the Dyson equations in 
      Ref.~\cite{Szczepaniak:2001rg,Szczepaniak:2003ve} and is referred to as model-1. The latter has  been advocated in 
       Ref.~\cite{Schleifenbaum:2006bq,Epple:2006hv} as the solution of the IR limit of the 
    Dyson equations which include the curvature but set the Coulomb form factor $f(p)=1$ and is referred to as model-2. These gap functions are shown in Fig.~\ref{omega-fig}. 
The gluelump spectrum for these two gap functions is displayed in Fig.~\ref{lattice_and_omega1and2}  together with the lattice results from Ref.~\cite{Bali:2003jq}.
The vacuum wave functional built from these gap functions systematically  overestimates the masses of the gluelumps by approximately $ 0.7\mbox{GeV}$. The inversion of the levels, due to the three-body interaction,   
 is clearly visible with $1^-$ above $1^+$ and $2^-$ near but still above $2^+$. In the next step we varied $\omega$ in the attempt to bring the $1^+$ state to be in agreement with lattice mass. We have found that the over mass scale is, as expected,  mostly sensitive to the value of $\omega(p)$ in the region $ 0.5 \mbox{ GeV} \lsim p \lsim 1.0\mbox{ GeV}$. We thus chose a new  functional form  for the gap function  which reduced $\omega$ in this momentum region. This is referred to model-$3$ in Fig.~\ref{omega-fig}.  The corresponding gluelump spectrum is shown in Fig.~\ref{lattice_and_omega3} where we show the  result of the calculation without the three-body interaction (dashed lines) and the full spectrum (solid lines). As discussed above the  two-body potential leads to the inverted spectrum with $1^-$ below  $1^+$ and $2^+$ below $2^-$ while the three-body interaction is operative for the natural parity states only, bringing the Coulomb gauge spectrum in  agreement with the lattice spectrum. 

\section{Summary and Outlook} 
The quenched gluelump state is the  simplest system that can be used to define a physical state of a gluon.  Within the framework of Coulomb-gauge QCD with a non-perturbative ansatz for the vacuum wave functional gluons correspond to quasi-particle excitations and low lying  gluelumps are states of a single quasi-gluon whose color is neutralized by an external static source. The possible quantum numbers of such states match with those found in lattice computations. Furthermore, ordering of these levels is reproduced, and it turns out to be non-trivially related to the non-abelian structure of the Coulomb  interaction in the Coulomb-gauge which leads to new (three-body) interaction in the Schr\"odinger  equation for the gluelump energies. The shape  of the gap function in the low ($p \lsim 1.0\mbox{ GeV}$) is primary relevant for setting the energy scale of the gluelumps. Because of the ambiguities in the renormalization program of variational Coulomb gauge calculations originating from the truncation of the Fock space, these  findings can be used to further constraint the Dyson equations of the mean field Coulomb gauge calculations.

\section{Acknowledgment}
We would like to thank  H.~Reinhardt and W.~Schleifenbaum  for  continuing discussions of the Coulomb gauge QCD.  This work was supported in part by the US Department of Energy grant under 
contract DE-FG0287ER40365. APS also thanks the  I.N.F.N and University of Genova for hospitality  during preparation of this work.

\end{document}